\documentclass[11pt]{article}
\usepackage{moriond,epsfig,amsmath}

\def\MyPhoto{\psfig{figure=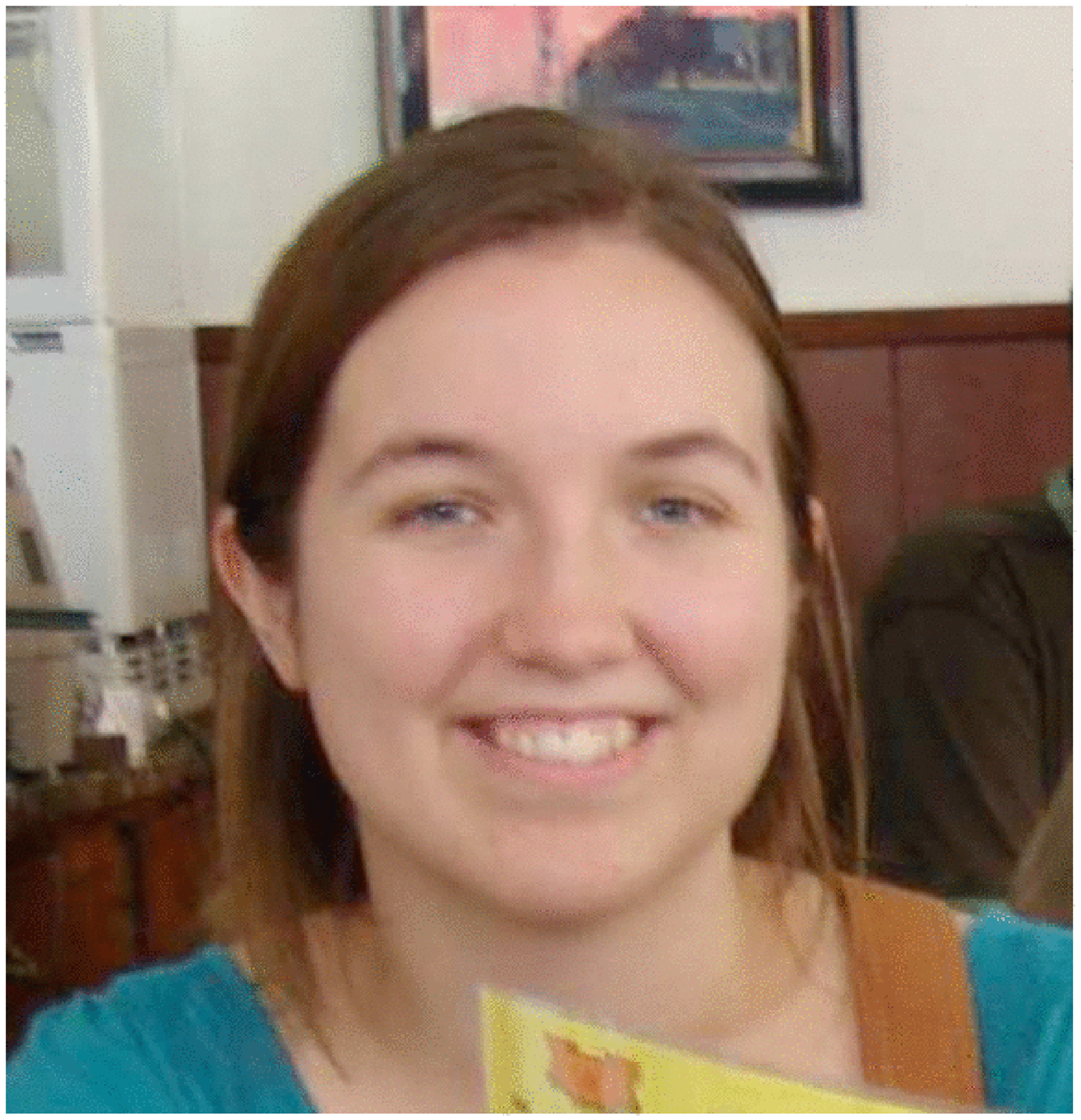,width=35mm}}

\def\PRD{{\em Phys. Rev.} D}
\def\nat{{\em Nature} (London)}
\def\apj{{\em Astrophys. J.}}
\def\aap{{\em Astron. Astrophys.}}

\begin{document}
\vspace*{4cm}
\title{DARK MATTER BOUND TO THE SOLAR SYSTEM: CONSEQUENCES FOR ANNIHILATION SEARCHES}

\author{ ANNIKA H.~G. PETER }

\address{California Institute of Technology, MS 249-17, Pasadena, CA 91125}

\maketitle
\abstracts{One method to search for particle dark matter is to hunt down its annihilation products. In the Solar System, three potential types of signals of annihilation have been identified: neutrinos and $\gamma$-rays from the Sun, and neutrinos from the Earth.  Each of these signals depends sensitively on the orbital evolution of dark matter once it becomes bound to the Solar System.  I will review progress on characterizing these signals based on recent improvements in the determination of the properties of the bound dark matter population.
}

\section{Introduction}\label{sec:intro}
Many lines of evidence point to the existence of dark matter in the universe, but its composition remains a mystery \cite{jungman1996}.  Although other particle physics (e.g., the axion \cite{weinberg1978}) and astrophysical (e.g., primordial black holes \cite{carr2005}) candidates exist, by far the most popular candidate for dark matter is one or more species of weakly interacting massive particle (WIMP).  Many proposed extensions to the Standard Model naturally produce such particles \cite{ellis1984}.

Numerous experimental methods are being used to determine the nature of dark matter.  Direct detection experiments are designed to measure the energy of recoiling nuclei after they have been hit by astrophysical WIMPs \cite{goodman1985}.  Byproducts from WIMP annihilation in dark matter concentrations throughout the Galaxy may be probed with cosmic ray experiments \cite{adriani2009}.  

The two brightest sources of annihilation products in the Solar System should be the Sun\cite{press1985} and Earth \cite{freese1986}.  WIMPs generically have small interaction cross sections with nuclei, and as such can be captured and settle into dense cores in the potential wells of celestial bodies.  Thus, WIMP densities $n$ in these bodies can be many orders of magnitude above the local Galactic density, which is important since the annihilation rate per unit volume $d\Gamma/dV \propto n^2$.  WIMP annihilation in the Earth may be detectable since it is a nearby source, while the Sun's far deeper potential well (and a correspondingly higher WIMP capture rate) compensates for its much greater distance from terrestrial detectors.

There are three possible signatures of WIMP annihilation in these bodies.  The only annihilation products which may escape from the interiors of the Sun and Earth are neutrinos.  The muon neutrinos may be observed in terrestrial neutrino telescopes (e.g., IceCube \cite{abbasi2009}, Antares \cite{hernandez2009}) via the Cerenkov radiation of neutrino-induced muons in and around the detector volume.  These signals may be distinguished from backgrounds by their directionality and energy spectrum.  Strausz \cite{strausz1999a} suggested that WIMPs annihilating just outside of the Sun may be visible in $\gamma$-rays (``near-solar $\gamma$-rays'').  Even though the annihilation rate is much lower outside the Sun than in its core, Strausz suggested that since the Sun is expected to emit few photons at high energies during its evolution, the backgrounds to the near-solar $\gamma$-ray signal would not be severe.  

In the next sections, I will describe the standard assumptions used in the calculations of these signals, and show how the calculations change when the details of the evolution of WIMPs bound to the Solar System are included.  

\section{Neutrinos from the Sun}\label{sec:nusun}
The upper limit on the neutrino flux from WIMP annihilation in the Sun \cite{abbasi2009,desai2004} currently places the tightest constraint on the WIMP-proton spin-dependent elastic scattering cross section $\sigma_p^{\mathrm{SD}}$.  The IceCube experiment is expected to have at least an order of magnitude better sensitivity once complete \cite{abbasi2009}, as will the proposed KM3NeT \cite{km3net2008}.  These experiments are projected to probe significant parts of the beyond Standard Model phase space, although this statement depends critically on the assumptions used to calculate event rates.

Schematically, the event rate of neutrino-induced muons in neutrino telescopes can be described by
\begin{eqnarray}
	\dot{N}_\mu \propto \Gamma \times \left(\hbox{neutrino physics}\right) \times \left(\hbox{detector properties}\right),
\end{eqnarray}
where $\Gamma$ is the total annihilation rate of WIMPs in the Sun and ``neutrino physics'' encompasses details on the annihilation branching fractions, propagation through the Sun and to the telescope, and interactions near the detector.  The standard calculation of the annihilation rate includes the following assumptions: (1) The thermalization is nearly instantaneous. (2) All WIMPs that are captured in the Sun thermalize. (3) The local Galactic WIMP distribution function is spatially smooth, and with velocities distributed as Maxwellian in an inertial Galactocentric frame (and with a one-dimensional velocity dispersion $\sigma = 155\hbox{ km s}^{-1}$).  I will touch on point (3) in Sec. \ref{sec:dd}, but describe the problems with the first two assumptions here.

Assumption (1) is violated due to the finite optical depth in the Sun to WIMPs, $\tau$, and the finite energy a WIMP may lose to a solar nucleus in each encounter, $Q$.  Since the Sun must be optically thin to WIMPs (deduced from existing constraints on the elastic scattering cross sections), WIMPs may make many passages through the Sun before rescattering, with the typical time between scatterings scaling as $t_r \sim P_\chi/\tau$, where $P_\chi$ is the WIMP orbital period.  In addition, it generically takes many scatters for a WIMP to thermalize because 
\begin{eqnarray}\label{eq:Q}
Q\sim m_A v_{esc}^2,
\end{eqnarray}
but the WIMP orbital energy is
\begin{eqnarray}
E \sim - m_\chi v_{esc}^2 (R_\odot/a), 
\end{eqnarray}
where $m_A$ is the mass of a solar nucleus with atomic number $A$, $m_\chi$ is the WIMP mass, $v_{esc}$ is the escape speed from the surface of the Sun, and $a$ is the semimajor axis.  A thermalized WIMP should have an energy
\begin{eqnarray}
	E_{therm}\ll - m_\chi v_{esc}^2.
\end{eqnarray}
As the WIMP mass increases, more scatters are required for the WIMP to sink to the center of the Sun.  Furthermore, as the WIMP mass increases, the median semimajor axis of the orbit to which a Galactic WIMP is captured increases, which implies that heavy WIMPs must lose more specific energy than lighter WIMPs to thermalize.  Thus, the median thermalization time for the solar captured WIMPs increases as the WIMP mass increases.  If spin-dependent (spin-independent, denoted SI) scattering dominates in the Sun, the median captured WIMP will require the age of the Solar System $t_\odot$ to thermalize if $\sigma_p^{SD} \approx 10^{-49}\hbox{ cm}^2$ ($\sigma_p^{SI} \approx 10^{-51}\hbox{ cm}^2$) if $m_\chi = 100$ GeV, but $\sigma_p^{SD} \approx 10^{-44}\hbox{ cm}^2$ ($\sigma_p^{SI} \approx 10^{-47}\hbox{ cm}^2$) if $m_\chi = 10$ TeV.  This is described in more detail in Peter \cite{peter2009b}.

Assumption (2) is violated due to the presence of planets in the Solar System, whose gravitational torques can affect orbits in two generic ways.  They can eject the WIMPs from the Solar System, meaning that those WIMPs will never thermalize; or they can alter the perihelia of the orbits so that WIMPs either pass less frequently through the Sun or only in the outskirts in the Sun where the optical depth is much lower.  In the latter case, thermalization may be significantly delayed, even beyond $t_\odot$.

I used a set of WIMP orbit simulations in a simplified solar system consisting of only the Sun and Jupiter (originally performed to determine the phase space density at the Earth of WIMPs bound to the solar system \cite{peter2009a}) to determine which effects dominated for a given initial semimajor axis.  I found that WIMPs initially scattered onto orbits with $a > 2.6$ AU were mostly ejected from the Solar System unless the thermalization time $t_t$ was less than the time required for the gravitational torque from Jupiter to lift the WIMP perihelion from the Sun ($\sim 1000$ years).  WIMPs with $1.5\hbox{ AU} < a < 2.6\hbox{ AU}$ had rescattering times $t_r \sim 300 P_\chi / \tau$.  This is longer than if the Sun were in isolation because secular and mean-motion resonances pull the WIMP perihelia out of the Sun for extended periods of time.  WIMPs with initial $a < 1.5$ AU had thermalization times largely unaffected by planetary torques.
  
\begin{figure}
	\psfig{figure=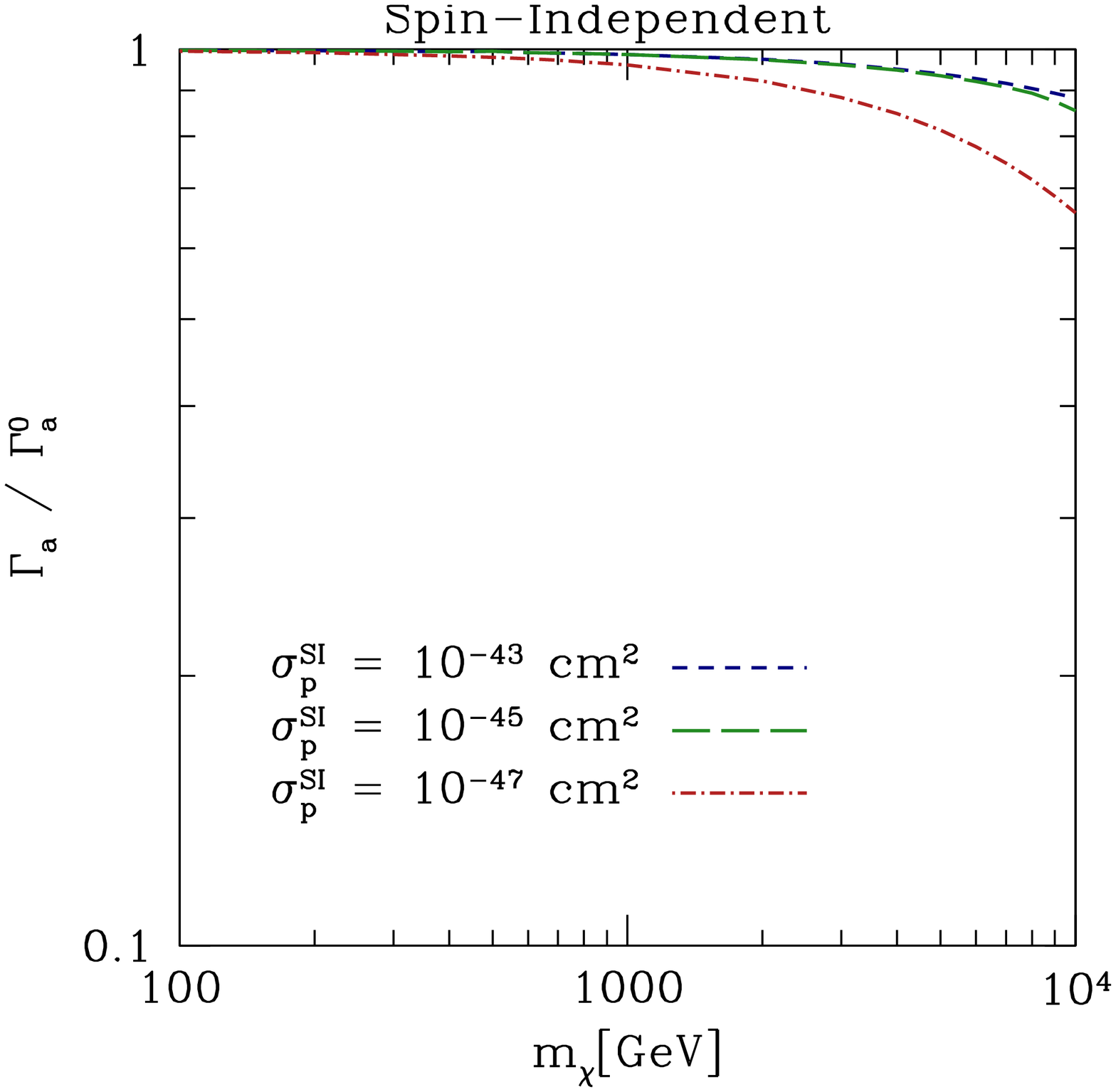,width=7cm} \psfig{figure=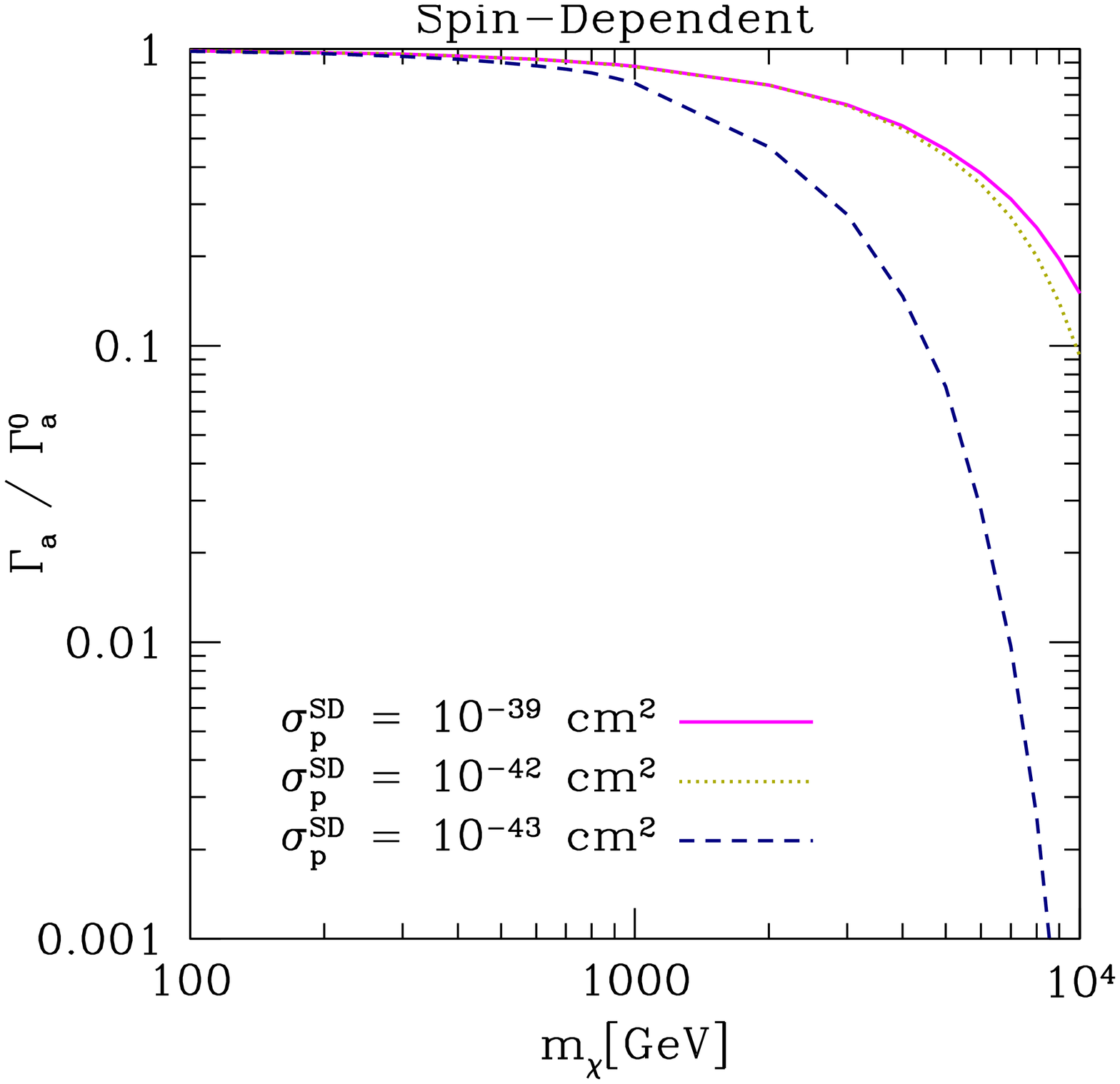,width=7cm}
	\caption{\label{fig:sun}Suppression of the annihilation rate due to the gravitational interactions of WIMPs with Jupiter in the cases that (\emph{left}) spin-independent and (\emph{right}) spin-dependent elastic scattering dominate in the Sun.  $\Gamma_a$ is the annihilation rate estimated from my simulations, and $\Gamma_a^0$ is the annihilation rate calculated using standard assumptions.  Figure from Peter $^{14}$.}
\end{figure}

The effects on the annihilation rate of violating assumptions (1) and (2) are shown in Fig. \ref{fig:sun}, in which I plot the ratio of the annihilation rate estimated from my simulations to the standard calculation as a function of WIMP mass.  Lines for several values of the elastic scattering cross section are show.  I fix the annihilation cross section to $\langle \sigma v\rangle = 3\times 10^{-26} \hbox{ cm}^3 \hbox{ s}^{-1}$, and use a conventional halo model for the local dark matter distribution.  In the left (right) panel, I show the suppression if spin-independent (spin-dependent) scattering dominates in the Sun.  For $\sigma_p^{SI} > 10^{-40}\hbox{ cm}^2$ ($\sigma_p^{SD} > 10^{-38}\hbox{ cm}^2$), there is no suppression because even the orbits with $a > 2.6\hbox{ AU}$ thermalize before gravitational torques from Jupiter can significantly affect the orbits.  WIMPs with $a < 2.6\hbox{ AU}$ also thermalize rapidly.  For the uppermost line in each panel of Fig. 1, the annihilation rate is suppressed due to the ejection of WIMPs with $a > 2.6\hbox{ AU}$.  However, WIMPs with $a < 2.6\hbox{ AU}$ thermalize on timescales less than $t_\odot$.  The suppression is greater at higher WIMP masses because an increasing fraction of WIMPs are initially captured onto long orbits, and is greater if spin-dependent scattering dominates in the Sun because the target nucleus on which the WIMPs scatter is much lighter than if spin-independent scattering dominates.  This reduces the typical energy loss per scatter (Eq. (\ref{eq:Q})).  The middle line in each panel shows suppression due to ejection, but at this point the capture and annihilation rates drop out of equilibrium in the Sun.  Whereas $\Gamma \propto C$, where $C$ is the capture rate of WIMPs in the Sun, if the rates are in equilibrium (with a coefficient that depends on whether WIMPs are self-annihilating or not), $\Gamma \propto C^2$ if the rates are far out of equilibrium.  In the lowest lines, the cross section is small enough that the thermalization time of the $1.5\hbox{ AU} < a < 2.6 \hbox{ AU}$ WIMPs is longer than $t_\odot$.  For cross sections much lower than these values, the thermalization time for WIMPs with $a < 1.5 \hbox{ AU}$ exceeds $t_\odot$, causing a near-total suppression of the annihilation rate.

In summary, the next generation of neutrino telescopes will have far lower sensitivity to high mass ($m_\chi > 1$ TeV) WIMPs than the standard annihilation calculation would suggest.

\section{Neutrinos from the Earth}\label{sec:nuearth}
The center of the Earth is an attractive target for WIMP searches due to its proximity to neutrino telescopes.  However, there are two complications in searching for annihilation in the Earth.  First, direct detection experiments constrain the spin-independent WIMP-proton cross section much more tightly than the spin-dependent cross section.  Since the Earth only has trace elements that may interact via spin-dependent channels with WIMPs (the dominant isotopes in the Earth, $^{56}$Fe, $^{16}$O, and $^{28}$Si, may only have spin-independent WIMP interactions), the capture rate is far more constrained than the capture rate of WIMPs in the Sun (in which the dominant species, hydrogen, may have spin-dependent interactions).

Second, the Earth's potential well is shallow---the escape speed from the center of the Earth is $v_{esc} \approx 15\hbox{ km s}^{-1}$ compared to the Sun's $\sim 1000$ km s$^{-1}$.  Typical halo WIMPs have speeds an order of magnitude higher than the escape speed from the Earth, which makes them kinematically difficult or impossible to capture in the Earth unless the WIMP mass is nearly the mass of one of the nuclear isotopes in the Earth \cite{gould1987}.  If $m_\chi > 400$ GeV, halo WIMPs are impossible to capture; the {\em only} WIMPs the Earth may capture are those already bound to the Solar System.

Thus, in order to predict event rates at neutrino telescopes (or to derive a WIMP annihilation rate from a signal), the Solar System's bound WIMP population must be characterized.  Using detailed balance arguments based on the gravitational scattering of WIMPs by planets, Gould \cite{gould1991} claimed that the annihilation rate could be accurately derived using the phase space density of WIMPs far outside the gravitational sphere of influence of the Solar System (the ``free space'' density).  More recently, Lundberg \& Edsj{\"o} \cite{lundberg2004} solved a diffusion equation for WIMPs in the Solar System, taking into account the loss of WIMPs in the Sun due to thermalization.  They found that the phase space density of bound WIMPs at the Earth was greatly reduced if the Sun were infinitely optically thick to WIMPs than if it were described as a gravitational point source.  Damour \& Krauss \cite{damour1999} described a population of WIMPs that could be captured by elastic scattering in the Sun, and survive for long times in the Solar System due to a secular resonance that forced the perihelion distance to the center of the Sun to oscillate.  If this population survived the lifetime of the Solar System, it could enhance the annihilation rate in the Earth by almost two orders of magnitude if $60\hbox{ GeV} < m_\chi < 130 \hbox{ GeV}\,$\cite{bergstrom1999}.  Annihilation rates are usually calculated using the free space distribution function, the purely unbound WIMP distribution function (which may be calculated at the Earth using Liouville's theorem and Galilean transformations), or Lundberg \& Edsj{\"o}'s phase space density.

\begin{figure}
\begin{center}
	\psfig{figure=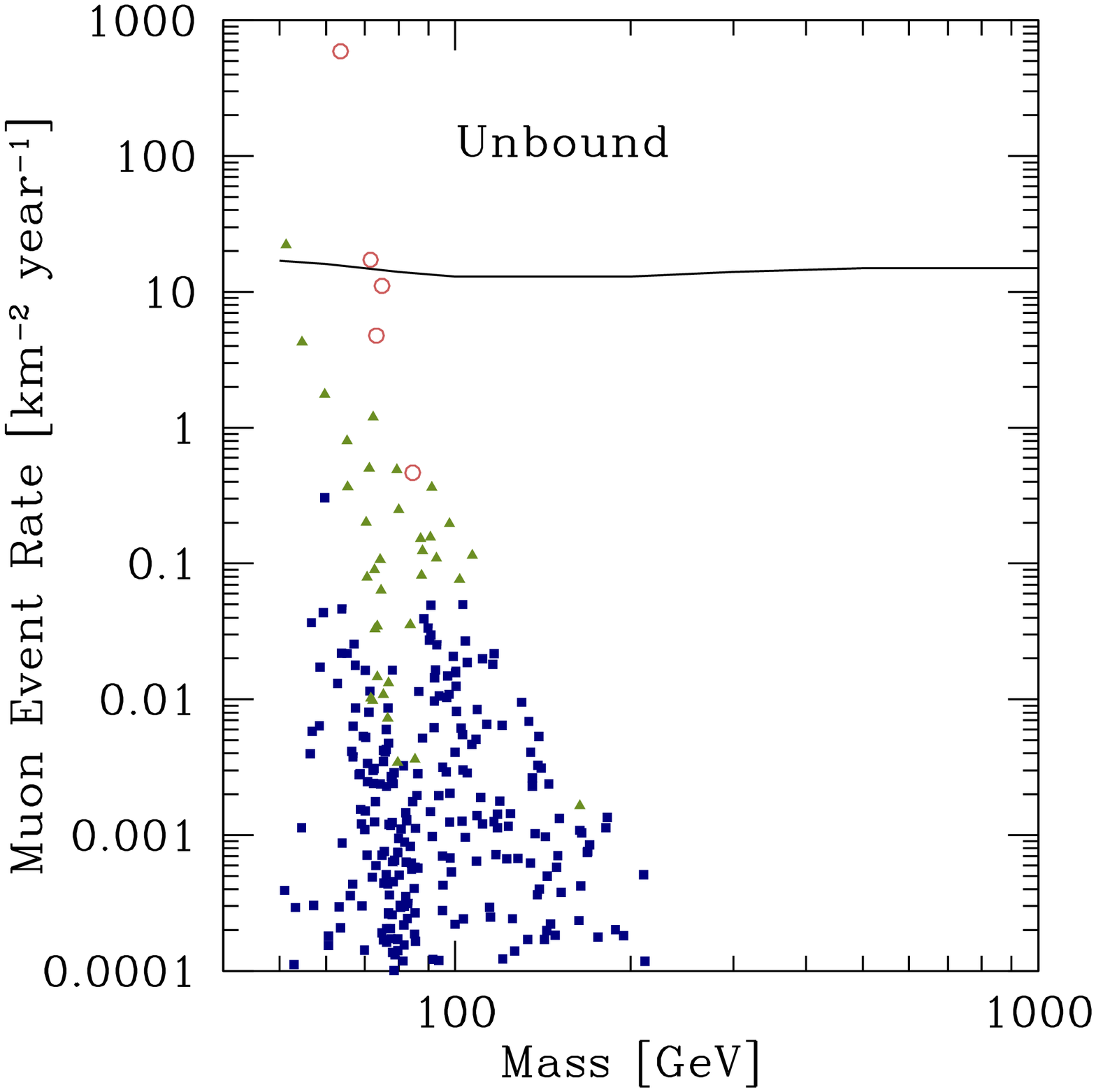,width=7cm} \psfig{figure=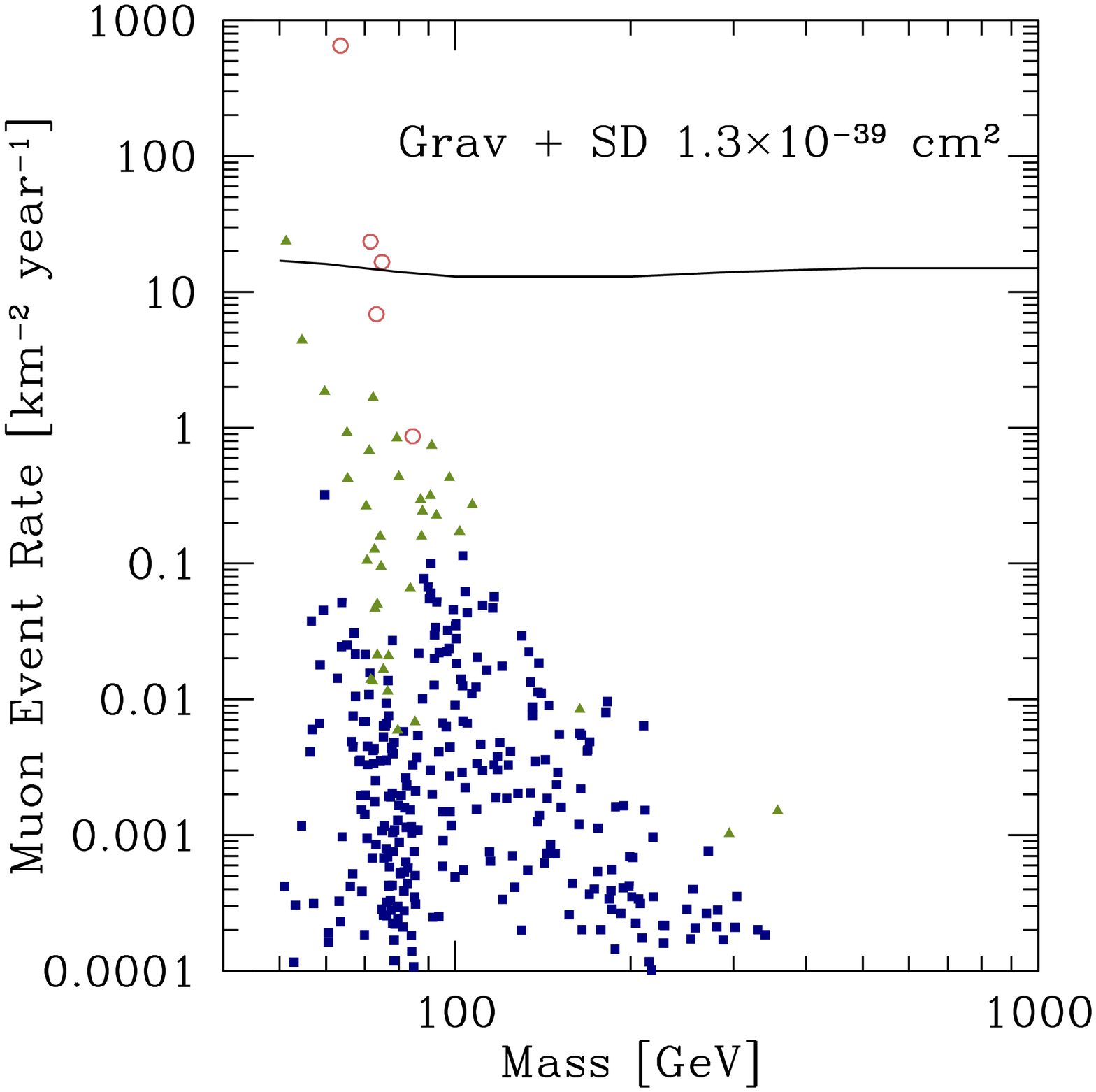,width=7cm}
\end{center}
	\caption{\label{fig:earth}Neutrino-induced muon fluxes from WIMP annihilation in the Earth above an energy threshold $E_\mu^{th} = 1$ GeV for water/ice neutrino telescopes for (\emph{left}) unbound halo WIMPs and (\emph{right}) the maximum expected from both bound and unbound dark matter, in scans of the MSSM.  The solid line represents an optimistic detection threshold for the IceCube telescope $^{18}$.  The open circles represent neutralino models with spin-independent cross sections above the 2006 CDMS exclusion curve $^{22}$, the triangles represent models with cross sections between the 2006 and the present limits, and models marked with blue squares are consistent with current bounds on the elastic scattering cross sections $^{23}$.  Figures from Peter $^{21}$.}
\end{figure}

Dynamics in the Solar System are far more complicated than can be described using the diffusion or semi-analytic treatments used in the work cited above to estimate the bound WIMP population.  The only way to truly determine the bound WIMP density in the Solar System is to simulate a statistically significant number of orbits.  As a first step, I simulated bound WIMP orbits in a solar system consisting only of the Sun and Jupiter.  I simulated both orbits bound to the solar system by elastic scattering in the Sun \cite{peter2009a} and WIMPs that are captured by gravitational interactions with Jupiter \cite{peter2009c}.  I found that the WIMPs bound by the first method have shorter lifetimes than assumed by Damour \& Krauss, and thus have a significantly lower density.  WIMPs that are gravitationally captured to the Solar System (the population explored by Gould and Lundberg \& Edsj{\"o}) have a slightly lower density than found by Lundberg \& Edsj{\"o}.

To estimate the neutrino-induced muon flux at the surface of the Earth, I assumed that the WIMP was the neutralino in the minimal supersymmetric standard model (MSSM), and used the DarkSUSY software package \cite{gondolo2004} to make scans of MSSM parameter space.  The results are shown in Fig. \ref{fig:earth}.  Fluxes due to unbound Galactic WIMPs only are shown on the left; those due to unbound WIMPs and the maximum possible bound WIMP density are on the right.  The solid line indicates an optimistic detection threshold for the IceCube neutrino telescope.  The squares mark models that are still allowed by direct detection experiments.  I find that the bound WIMPs only marginally increase the fluxes at the Earth, and that allowed models lie well below threshold.

There are two caveats to this pessimistic conclusion.  First, I have only shown the flux of through-going muons; since IceCube is large compared to typical muon path lengths through ice, there could be a large number of contained events.  However, this is likely to increase the event rate above the through-going event rate only by factors of several to about a factor of ten.  This is still insufficient to boost the muon flux above threshold.

Second, the regions of WIMP phase space that are most accessible to capture are not accessible in the simplified solar system.  This part of phase space can only be populated by close encounters between WIMPs and the inner planets.  Therefore, more sophisticated simulations are required to accurately predict the signal of WIMP annihilation in the Earth.

\subsection{Everything Is Not Lost: Dark Matter Disks and Detection}\label{sec:dd}
Although the prospects for detecting WIMP annihilation in the Earth are grim, these conclusions are drawn based on fairly strict assumptions on how dark matter is distributed in the Galaxy, and what its phase space structure is.  It is almost always assumed that the dark matter is distributed almost smoothly in a nearly spherically symmetric Navarro-Frenk-White halo\cite{navarro1997} about the Galactic center.  Moreover, it is assumed that the halo is at rest in an inertial Galactocentric frame, and that the velocity distribution can be described locally by a Maxwellian with a velocity dispersion $\sigma = v_\odot/\sqrt{2}$, where the stellar disk has a rotation speed of $v_\odot = 220 \hbox{ km s}^{-1}$.  These assumptions are based on N-body simulations of dark matter in roughly Milky Way-sized halos.  

However, the Sun lies in part of the Galaxy that is dominated by luminous matter, not dark matter.  Any local dark matter parameters that are derived from simulations should be derived from simulations that include luminous matter, too.  Recently, Read et al. \cite{read2008,read2009} have analyzed simulations of disk galaxies in Milky Way-mass halos, in both idealized and cosmological contexts.  They find that, in general, massive satellite galaxies are dragged into the spiral galaxy disk plane and dissolved, yielding a ``dark disk'' of dark matter.  This component is in addition to the conventional, more spherical dark matter halo.  The properties of the dark disk depend sensitively on the accretion history of the galaxy, but the ``median'' galaxy in their samples had a dark disk with a rotation velocity lag with respect to the Sun of $\sim 50\hbox{ km s}^{-1}$, and a velocity dispersion of $\sigma \approx 50\hbox{ km s}^{-1}$, and a mass density in the mid-plane similar to that of the halo.

This dark disk is important for WIMP detection because the typical WIMP speed with respect to the Solar System is much lower than for WIMPs in the halo.  The capture probability in the Solar System increases dramatically with decreasing relative speed.  In work lead by Tobias Bruch \cite{bruch2009}, we found the consequence was that the neutrino flux from WIMP annihilation in the Earth could be increased by two or three {\em orders of magnitude} assuming the median dark disk, and the neutrino flux from the Sun could be boosted approximately an order of magnitude.  Even for the least significant dark disk found in the Read et al. samples, the neutrino flux from the Earth and the Sun could be enhanced by a factor of order unity.  Therefore, the dark disk may boost the neutrino flux for WIMP models consistent with experiments (farther) above the detection threshold for IceCube.

There are two major sources of uncertainty in the dark disk enhancement of the annihilation rate.  First, the properties of the dark disk are highly uncertain.  The dark disk is likely to be difficult to probe with stellar dynamics, and any stellar debris from satellite destruction in the disk will also be difficult to disentangle from other stellar populations.  Dark matter experiments are the most sensitive probes of the dark disk.  The energy spectrum of events in direct detection experiments should yield some constraints on dark disk properties \cite{bruch2009a}, but such constraints will take years (and an actual detection of dark matter in more than one experiment!) to materialize. 

Second, whether a high mass WIMP is detectable or not in neutrinos from the Earth depends sensitively on the phase space density of WIMPs bound to the Solar System.  This is illustrated in Fig. 3 of Bruch et al.\cite{bruch2009}, which shows that for, e.g., a 400 GeV WIMP, the annihilation rate is about three orders of magnitude higher for Gould's free space phase density than for the phase densities found in my simulations.  Therefore, more sophisticated versions of my simulations, including more planets and more realistic planet orbits, will be required to determine the bound dark matter phase space density to sufficient precision to estimate WIMP parameters from neutrino telescope data.

\section{Near-Solar $\gamma$-Rays}\label{sec:gammasun}
Strausz \cite{strausz1999a} suggested that WIMP annihilations occuring just outside the Sun may be visible in $\gamma$-rays.  Although the dark matter density is far lower outside the Sun than at its center, it is higher right outside the Sun than far outside the Sun's gravitational sphere of influence.  This is due to several effects.  First, gravitational focusing increases the unbound WIMP density deep in the Sun's potential well.  Second, in the absence of kinematic suppression, WIMPs are captured onto bound orbits with a semimajor axis distribution $d\log N/d\log a = -1$.  Third, massive WIMPs (much heavier than the typical solar nucleus) lose energy during each encounter with a solar nucleus.  WIMPs initially captured onto barely bound orbits will spend time just outside the Sun as they thermalize.  Moreover, the Sun does not produce $\gamma$-rays at its surface as a result of stellar evolution.  Strausz suggested the Milagro, an air shower array used to detect cosmic rays, should be able to detect these near-solar $\gamma$-rays.  However, it was only able to place an upper limit to the near-solar $\gamma$-ray flux \cite{atkins2004}.  

Recent work suggests that the WIMP-generated $\gamma$-ray signal should be virtually undetectable.  Hooper \cite{hooper2001} and Sivertsson \& Edsj{\"o} \cite{sivertsson2009} have redone Strausz's calculations, and find that the $\gamma$-ray flux should be many orders of magnitude lower than Strausz's predictions.  Using Monte Carlo realizations of the thermalization process, Sivertsson \& Edsj{\"o} predict a flux of  $\sim 10^{-7} \hbox{ km}^{-2} \hbox{ yr}^{-1}$ if $\sigma_p^{SD} = 10^{-39}\hbox{ cm}^2$ and $m_\chi = 1$ TeV.  Using my own simulations, I have also estimated the expected $\gamma$-ray flux, and find similar results, although the flux is suppressed even further for high mass WIMPs due to the gravitational effects described in Sec. \ref{sec:nusun}.  It is not clear what the error in Strausz's calculation is, but current calculations suggest that the flux of near-solar $\gamma$-rays at the Earth should be tiny.

In addition, the Sun is more luminous $\gamma$-rays than initially postulated by Strausz.  Cosmic rays protons can produce pion showers in the chromosphere of the Sun \cite{seckel1991}, which produce $\gamma$-rays by $\pi^0 \rightarrow \gamma\gamma$.  Seckel et al. \cite{seckel1991} estimated that the total $\gamma$-ray flux from this process should be $\sim 10^{10} \hbox{ km}^{-2} \hbox{ yr}^{-1}$ ($\sim 10^{-7} \hbox{ cm}^{-2} \hbox{ s}^{-1}$) above $1$ MeV, and about an order of magnitude less above 1 GeV.  

Inverse-Compton scattering of solar photons on cosmic ray electrons can also produce a significant halo of $\gamma$-rays around the Sun.  Orlando \& Strong \cite{orlando2008} find a detection of this signal in EGRET data, a flux of $\sim 3 \times 10^{-7} \hbox{ cm}^{-2} \hbox{ s}^{-1}$ in the $100-300$ MeV energy band, which is expected to be stronger than the cosmic ray proton-induced $\gamma$-ray flux by factors of several.  The Fermi Gamma-ray Space Telescope (\emph{Fermi}) should be exquisitely sensitive to these cosmic ray-induced $\gamma$-rays from the Sun, and the signal should completely drown out any WIMP-induced signal from the Sun.

\section{Conclusion}\label{sec:conclusion}
In this paper, I have described recent updates to the estimates of the solar and terrestrial neutrino and near-solar $\gamma$-ray fluxes from WIMP annihilation at the centers of the Sun and Earth due to new estimates of the orbital evolution of WIMPs in the Solar System.  Some of the effects I described may change the event rates by orders of magnitude.  I would like to emphasize that in order to understand the particle physics of WIMPs using new astrophysical data sets (whether it be neutrinos or $\gamma$-rays from Solar System sources or $\gamma$-rays from the Galactic center), it is necessary to understand the astrophysical properties of WIMPs: their distribution throughout the Galaxy, on both Solar System and $\sim$ kiloparsec scales; and the velocity structure.  While there has been a lot of progress in constraining the astrophysical properties of WIMPs, there are still many large uncertainties in these properties.


\section*{Acknowledgments}
This work was supported by NASA grants No. NNG04GL47G and No. NNX08AH24G and by the Gordon and Betty Moore Foundation.

\end{document}